\documentclass[twocolumn,showpacs,preprintnumbers,amsmath,amsfonts,amssymb,floatfix,aps,pra,superscriptaddress,nopacs]{revtex4}

\usepackage{color}
\usepackage{amsmath}
\usepackage{amssymb}
\usepackage{graphicx}
\usepackage{bm}
\usepackage{hyperref}
\usepackage{float}

\def\be{\begin{equation}}
\def\ee{\end{equation}}

\usepackage{cancel}

\newcommand{\ds}{\displaystyle}

\def\Blue#1{{\color{black}{#1}}}

\begin{document}

\title{Nonlinear dynamics of coupled superfluids}

\author{S. Laurent}%
\affiliation{Laboratoire Kastler Brossel, ENS-Universit\'e PSL, CNRS, Sorbonne Universit\'e, Coll\`ege de France.
}%

\author{P. Parnaudeau}%
\affiliation{Institut Pprime, CNRS,
Universit{\'e} de Poitiers - ISAE-ENSMA - UPR 3346,\\
11 Boulevard Marie et Pierre Curie,
86962 Futuroscope Chasseneuil Cedex, France
}%

\author{F. Chevy}%
\email{Corresponding author: chevy@lkb.ens.fr}
\affiliation{Laboratoire Kastler Brossel, ENS-Universit\'e PSL, CNRS, Sorbonne Universit\'e, Coll\`ege de France.
}%

\author{I. Danaila}%
\affiliation{Laboratoire de Math{\'e}matiques Rapha{\"e}l Salem,
	Universit{\'e} de Rouen Normandie, CNRS UMR 6085, 76801 Saint-{\'E}tienne-du-Rouvray, France
}%

\date{\today}

\begin{abstract}
Following recent experiments on ultracold dual superflows, we model in this work the dynamics of two harmonically trapped counterflowing superfluids. Using complementary analytical and numerical approaches, we study the shedding of elementary excitations triggered by the relative motion of the two species. We exhibit two different excitation mechanisms leading to distinct threshold velocities for the onset of dissipation: in addition to the parametric pair production present in homogeneous, galilean-invariant systems, we show that non-uniform motion and density inhomogeneities allow for a Landau-like decay mechanism where single excitations are produced.
\end{abstract}

\maketitle

\section{Introduction}

In recent years, progress in manipulation of ultracold boson/boson \cite{myatt1997production} or boson/fermion mixtures \cite{Ferrier2014Mixture,ikemachi2017all,Roy17two,Yao16Observation} has led to the observation  of dual-superfluid gases. In these systems, the interplay between the two species leads to a rich physics, like the creation of spin domains \cite{stenger1998spin}, dark-bright solitons \cite{Busch2001Dark,tylutki2016dark} or ``liquid" droplets \cite{petrov2015quantum,cabrera2017quantum}.  Dual-superfluid mixtures play a fundamental role going beyond the field of atomic physics and quantum gases. In neutron stars,
 kaon condensates are believed to coexist with neutron-proton superfluids \cite{ellis1995kaon} while among the liquid Helium community the quest for a dual $^3$He/$^4$He dual superfluid mixture has been considered as one of the holy grails of low-temperature physics  ever since the observation of superfluidity in $^3$He \cite{Rysti2012effective}.

An intriguing possibility is the existence of superfluid counterflows where the two fluids move with different velocities. This  multiple-fluid hydrodynamics  was considered first theoretically for Helium mixtures \cite{volovik1975theory,andreev1976three,nespolo2017andreev} and was observed experimentally in ultracold boson/boson \cite{maddaloni2000collective,smerzi2003macroscopic,hamner2011generation} and fermion/boson mixtures \cite{delehaye2015critical,Yao16Observation,wu2018coupled}. These experiments  raised the question of the critical relative velocity above which dissipation arises and suggested that superfluid counterflows were unstable against generation of pairs of excitations, a mechanism providing a Galilean-invariant generalization of Landau's celebrated scenario \cite{castin2014landau,zheng2014quasiparticle,abad2014counter}.

In the experiments reported in \cite{delehaye2015critical}, the counterflow was obtained by releasing a dual Bose/Fermi superfluid of $^6$Li and $^7$Li in a harmonic trap. Due to the mass difference between the two species, the motion of the two superfluid get out of phase after a few oscillation periods. This creates a relative motion between the two components and dissipation was only observed above a critical velocity confirming the superfluid nature of the counterflow. However,  the harmonic trapping confining the atoms breaks translational and Galilean invariance assumptions underpinning simple dissipation mechanisms.  In the present work, we explore theoretically the stability of the counterflow generated by two harmonically trapped superfluids. We first consider the case of two weakly interacting Bose-Einstein condensates described using coupled Gross-Pitaevskii equations. These equations are solved numerically and the density profiles are characterized using a Principal Component Analysis (PCA) scheme \cite{jolliffe2002introduction} that allows for a model-free identification of the modes triggered by the relative motion of the two clouds.  In the second part of the paper we investigate the mode-coupling mechanism using an analytical hydrodynamic approach. This approximation is valid only for long wavelength excitations but it is complementary to the numerical simulation, since it is applicable to  any kind of superfluid (bosons or fermions, weakly or strongly coupled, etc.). We show that the relative motion excites eigenmodes of the two clouds and we identify two different excitation mechanisms. First, a parametric process akin to \cite{castin2014landau} leads to the formation of pairs of excitations in both superfluids. Second,  the motion of the smaller cloud excites linearly the density profile of the larger one by the creation of single phonon-like modes, following a process similar to Landau's traditional scenario. We show in particular that this latter process is a direct consequence of the translation-symmetry breaking induced by the presence of the trap.

\section{Numerical simulations}

\subsection{Coupled Gross-Pitaevskii equations}

We first simulate numerically the counterflow of two harmonically trapped superfluids using a set of  coupled Gross-Pitaevskii equations describing  zero-temperature Bose-Einstein condensates (BECs) with mean-field interactions:
\begin{eqnarray}
\label{eq-GP_psi1}
\ds i \hbar  \frac{\partial \psi_1}{\partial t}=& \ds  \left[-\frac{\hbar^2}{2m_1}\nabla^2  + U(\bm r) +U_{{\rm mf},1}(\bm r)\right]\psi_1,&\\
\label{eq-GP_psi2}
\ds  i \hbar \frac{\partial \psi_2}{\partial t} =&\ds \left[-\frac{\hbar^2}{2m_2} \nabla^2  + U(\bm r) + U_{{\rm mf},2}(\bm r) \right]\psi_2,&
\end{eqnarray}
\Blue{with  $N_1$ and $N_2$ the number of particles of each species. The wavefunctions $\psi_{1}$ and $\psi_2$ are thus normalized to unity.} To reproduce the setting used in recent experiments \cite{Ferrier2014Mixture}, the two species are trapped using the same cigar-shaped harmonic external potential
\be
U(\bm r)=\frac{1}{2}m_2\left[\omega_\perp^2(x^2+y^2)+\omega_z^2z^2\right],
\ee
where $\omega_\perp$ and $\omega_z$ are  the radial and axial trapping frequencies, respectively. Finally $U_{{\rm mf}, 1,2}$ describe the mean-field interaction between the atoms and we have $U_{{\rm mf},i}=\sum_{j=1,2} N_j g_{ij}|\psi_j|^2$, where the $g_{ij}$ are the s-wave coupling constants satisfying the symmetry condition $g_{ij}=g_{ji}$.

\begin{figure}[!h]
	\includegraphics[width=0.8\columnwidth]{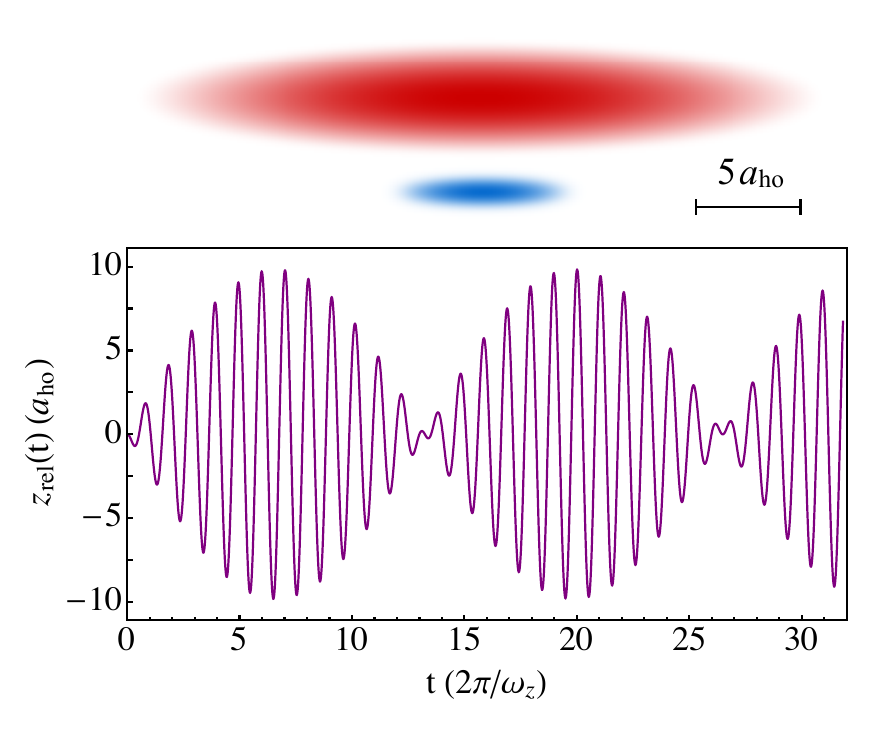}
	\caption{\Blue{Example of low amplitude oscillations of the relative position {$z_{rel}$} of centers of mass of the two superfluids. Case $b/a_{ho}=2$ and $g_{12}/g_{22}=0.031$. The small atomic mass difference $(m_1-m_2)/m_1\ll1$ combined with a large atom number ratio $N_2/N_1=30$ induce a strong amplitude modulation of the small condensate oscillations ($i=1$, in blue), while the oscillations of the large condensate  are mostly not affected ($i=2$, in red).}		
	}	
	\label{fig:Cdm}
\end{figure}

The parameters are chosen  to reproduce two important features of the experiment described in \cite{Ferrier2014Mixture}. First, we consider an atomic mass ratio of  $m_{1}/m_{2}=7/6$, such that the bare trapping frequencies of the two clouds are related via $\omega_{z,1}=\sqrt{6/7}\,\omega_{z}$. Thus, by displacing the clouds by the same distance $b$ along the $z$-axis, a relative motion between the clouds  progressively sets in and a periodic counterflow between the two superfluids is   created. In addition, the small atomic mass difference allows for a strong coherent energy exchange, as  observed in \cite{Ferrier2014Mixture}. Second, we chose very different atoms numbers and interspecies coupling constants to create a large shape asymmetry between the two clouds. The $i=1$ cloud is weakly interacting and small while the $i=2$ cloud  is strongly repulsive and is much broader (see density profiles in the top panel of Fig.\,\ref{fig:Cdm}).

We start the numerical simulation by generating the static density profiles of the two Bose-Einstein condensates. For this purpose, we numerically solve the coupled stationary Gross-Pitaevskii equations corresponding to (\ref{eq-GP_psi1})-(\ref{eq-GP_psi2}) using  an imaginary-time propagation method with Fourier spectral space accuracy \cite{BEC-baow-2004-Du,BEC-numm-2014-antoine-duboscq-JCP}.
Then we shift the two stationary profiles and let the system evolve in time. The real dynamics described by (\ref{eq-GP_psi1})-(\ref{eq-GP_psi2})  is computed using a second-order Strang splitting method \cite{BEC-review-2013-antoine-besse-bao}. Both stationary and real-time dynamics computations are performed using the state-of-the-art computational code GPS (Gross-Pitaevskii-Simulator) offering various modern numerical methods to solve GP equations on high-performance parallel computers \cite{BEC-parnaudeau}. A typical grid used for this study contained $1024\times 128\times 128$ computational points along the $z$ and $x, y$ axes, respectively.
To capture the oscillations of the system for long times, up to $400,000$ time steps were needed. Special care was devoted to the accuracy of the time and space numerical scheme, in order to conserve the mass and energy of the system during this long-time integration. More details on the numerical simulations are given in Appendix \ref{AppendixSim}). \Blue{Supplemental Material is provided with animations depicting the oscillations of the clouds for several run cases~\cite{SupMat}.}

\Blue{The influence of the coupling parameters was explored by simulating more than 30 cases, for values in the range $0 < g_{12}/g_{22} < 0.3$ and $2 \leq b/a_{ho} \leq 8$, with $a_{ho}=\sqrt{\hbar/m\omega_z}$.}

\Blue{A typical numerical result is displayed in Fig. \ref{fig:Cdm}.  The  response of the center of mass of the two clouds is illustrated by plotting their relative position {$z_{rel} =z_2(t)-z_1(t)$}.  For the case $g_{12}/g_{22}=0.031$, $b=2$, low amplitude oscillations of the clouds are observed and $z_{rel}$ beats
with the frequency difference $\omega_{z,1}-\omega_{z,2}$.}

\begin{figure}[!h]
	\includegraphics[width=0.8\columnwidth]{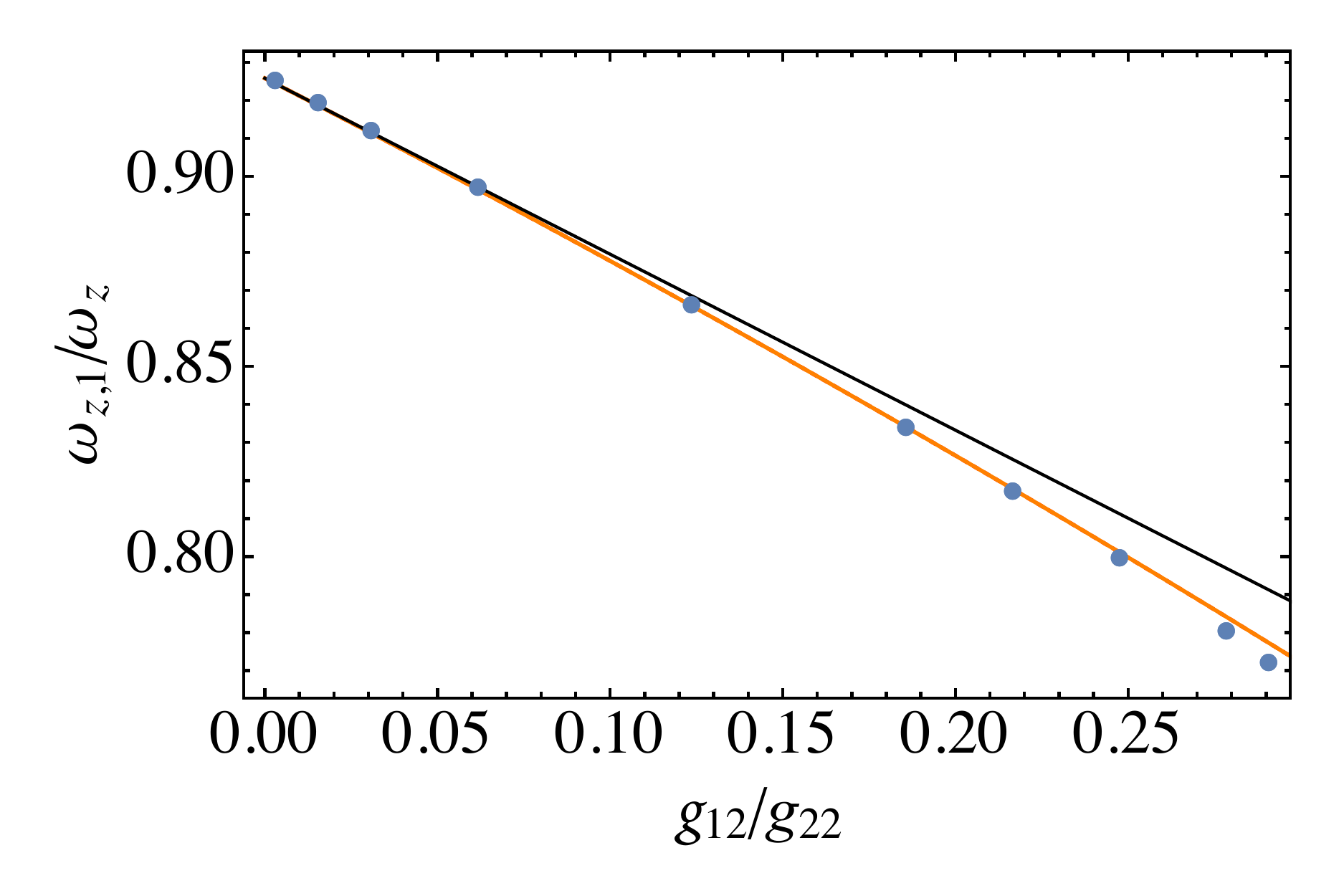}
	\caption{\Blue{Dipole-frequency shift ($\omega_{z,1}/\omega_{z}$) of the oscillations of the small cloud ($i=1$) for different interspecies coupling $g_{12}/g_{22}$:  numerical simulation (blue dots), theoretical prediction (black line) based on the perturbation theory Eq. (\ref{EqnDipoleShift}) and  coupled-oscillator prediction (orange line).}}	
	\label{fig:Cdm-2}
\end{figure}

\begin{figure*}
	\includegraphics[width=\textwidth]{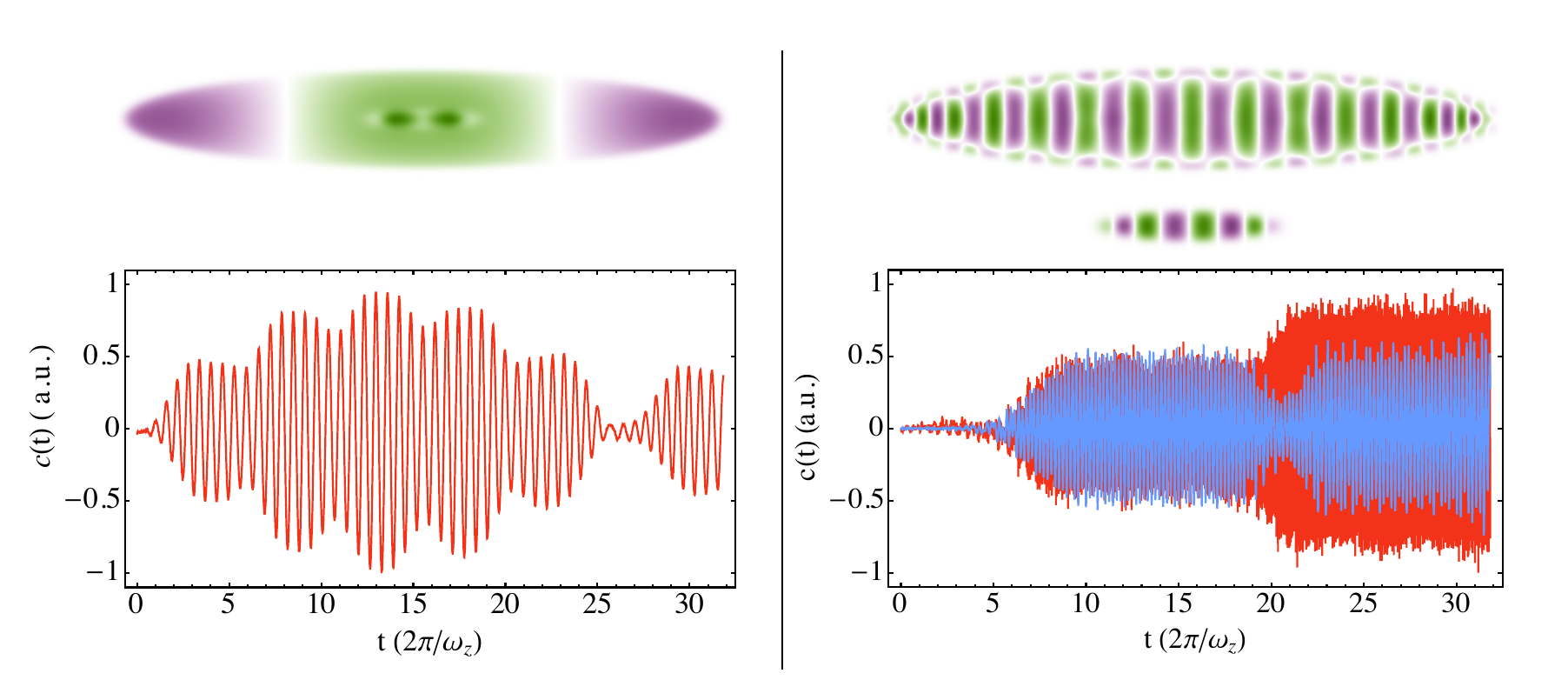}
	\caption{\Blue{Examples of the two nonlinear modes observed in simulations and analysed using the Principal Component Analysis (PCA). Space structure (upper panels) and time evolution of the amplitude of the modes (lower panels) extracted using the PCA.  Blue lines are for the small cloud ($i=1$) and red lines for the  large cloud ($i=2$). Linear forced mode (left, $g_{12}/g_{22}=0.291$, $b/a_{ho}=1$) and parametric modes (right, $g_{12}/g_{22}=0.003$, $b/a_{ho}=5$).  Animations depicting the oscillations of the clouds for these regimes are provided in the Supplemental Material ~\cite{SupMat}.}}
	\label{fig:param}
\end{figure*}

\subsection{Dipole-mode frequency shift}
\Blue{Numerical simulations similar to that displayed in Fig. \ref{fig:Cdm}, allowed us to follow the evolution of the oscillation frequency of the small cloud with the interspecies coupling constant $g_{12}/g_{22}$.  Figure \ref{fig:Cdm-2}  confirms the  shift of the dipole mode, as predicted by the sum-rule approach that was previously used in \cite{Ferrier2014Mixture} to measure the equation of state of an attractive Fermi gas in the BEC-BCS crossover or more recently in \cite{Roy17two} to measure the Lithium-Cesium scattering length and in \cite{wu2018coupled} in the case of LiK mixtures. In our case,} the sum-rule associated with the shift of the two trapping potentials predicts that the oscillation frequencies of the two clouds are the eigenvalues of the susceptibility matrix $\cal M$ defined by
\be
{\cal M}=
\left(
\begin{array}{cc}
m_1(1-\frac{N_1}{N_2}\chi_{12})&\sqrt{m_1 m_2}\sqrt{\frac{N_1}{N_2}}\chi_{12}\\
\sqrt{m_1 m_2}\sqrt{\frac{N_1}{N_2}}\chi_{12}&m_2(1-\chi_{12})
\end{array}
\right)
\ee
where $\chi_{12}=\frac{\partial z_1}{\partial b_2}$ is the displacement of the center of cloud $i=1$ after a shift by a distance $b_2$ of the potential trapping \Blue{of cloud} $i=2$. In the weak-coupling limit, this model predicts a shift $\delta\omega$ of the smaller cloud frequency given by
\be
\frac{\delta\omega_{1,z}}{\omega_{1,z}}\simeq-\frac{g_{12}}{2}\left(\frac{\partial\rho_2}{\partial\mu_2}\right)_0,
\label{EqnDipoleShift}
\ee
where $\rho$ and $\mu$ are the density and the chemical potential of the gas, respectively.

We observe in \Blue{Fig. (\ref{fig:Cdm-2})} that the simulation agrees with this asymptotic regime for a weak coupling. For intermediate coupling, we observe a slight departure of the simulation results with respect to Eq. (\ref{EqnDipoleShift}) but this discrepancy can be cured using the exact diagonalization of $\cal M$ \Blue{(orange line in Fig. \ref{fig:Cdm-2})}.

\subsection{Mode coupling}

To get further insight on the dynamics of the system, we analyze the density profiles of the two clouds by performing a Principal Component Analysis (PCA) \cite{jolliffe2002introduction}. This method allows us to identify the modes involved in the dynamics of system without any a priori assumption on their spacial structure \cite{dubessy2014imaging}. \Blue{When applied to cloud images (atomic density integrated along the $y$ axis), the PCA extracts a set of 4000 modes together
with their associated eigenvalues and their temporal evolutions. Only a limited number of modes play
a non-negligible role in the fluctuations (at least 99\% of the eigenvalues are $10^6$ times smaller than the largest
one). For each run, we typically limited our analysis to the 10-30 most populated modes given by the PCA. Among these modes, we could identify the following types of modes: parametric, linear forced, filtered dipole modes and harmonic modes.  The first two types of nonlinear modes are illustrated in Fig. \ref{fig:param} and additional modes are depicted in in the Supplemental Material ~\cite{SupMat}.}

\begin{figure}
\centerline{\includegraphics[width=\columnwidth]{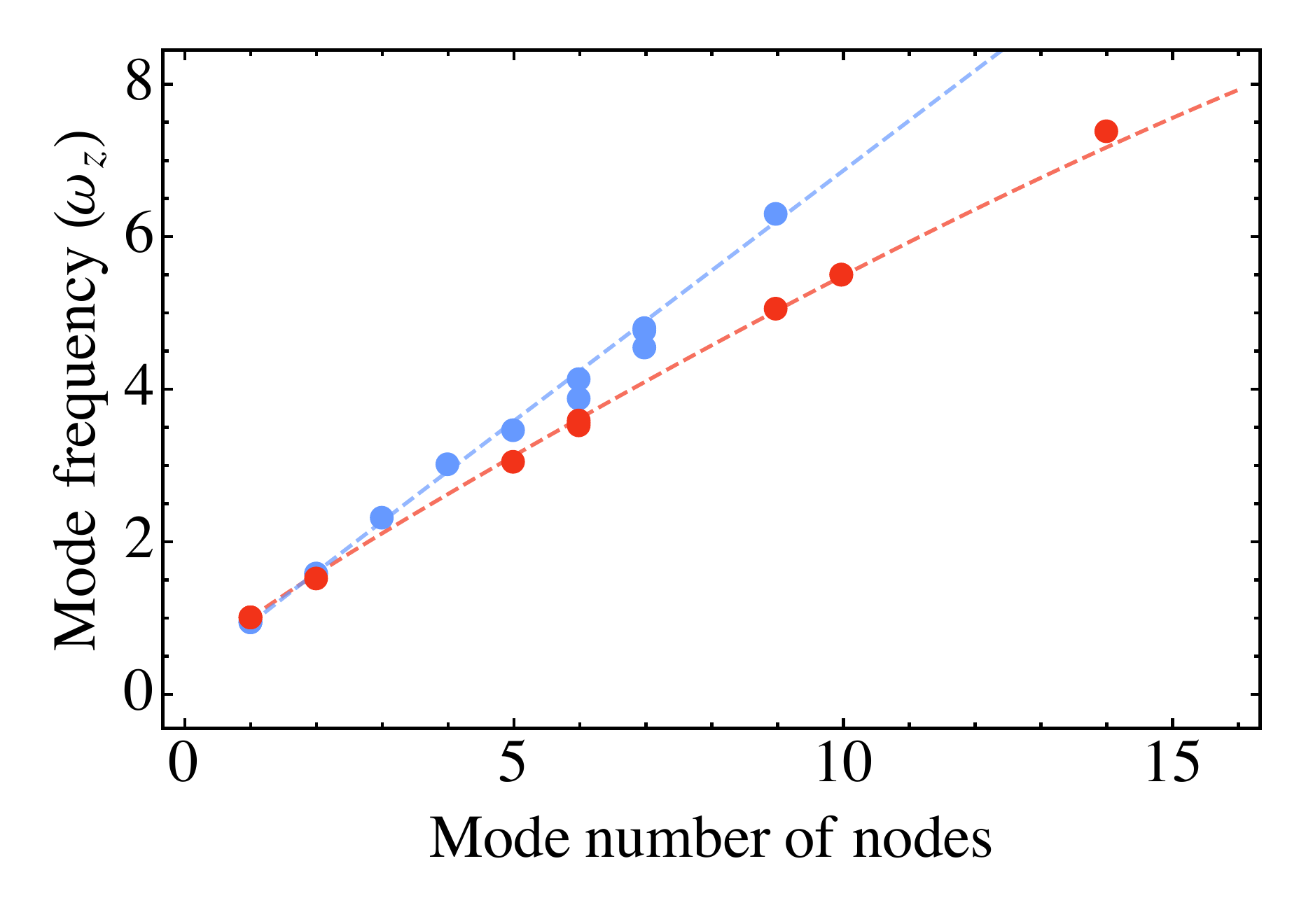}}
\centerline{\includegraphics[width=\columnwidth]{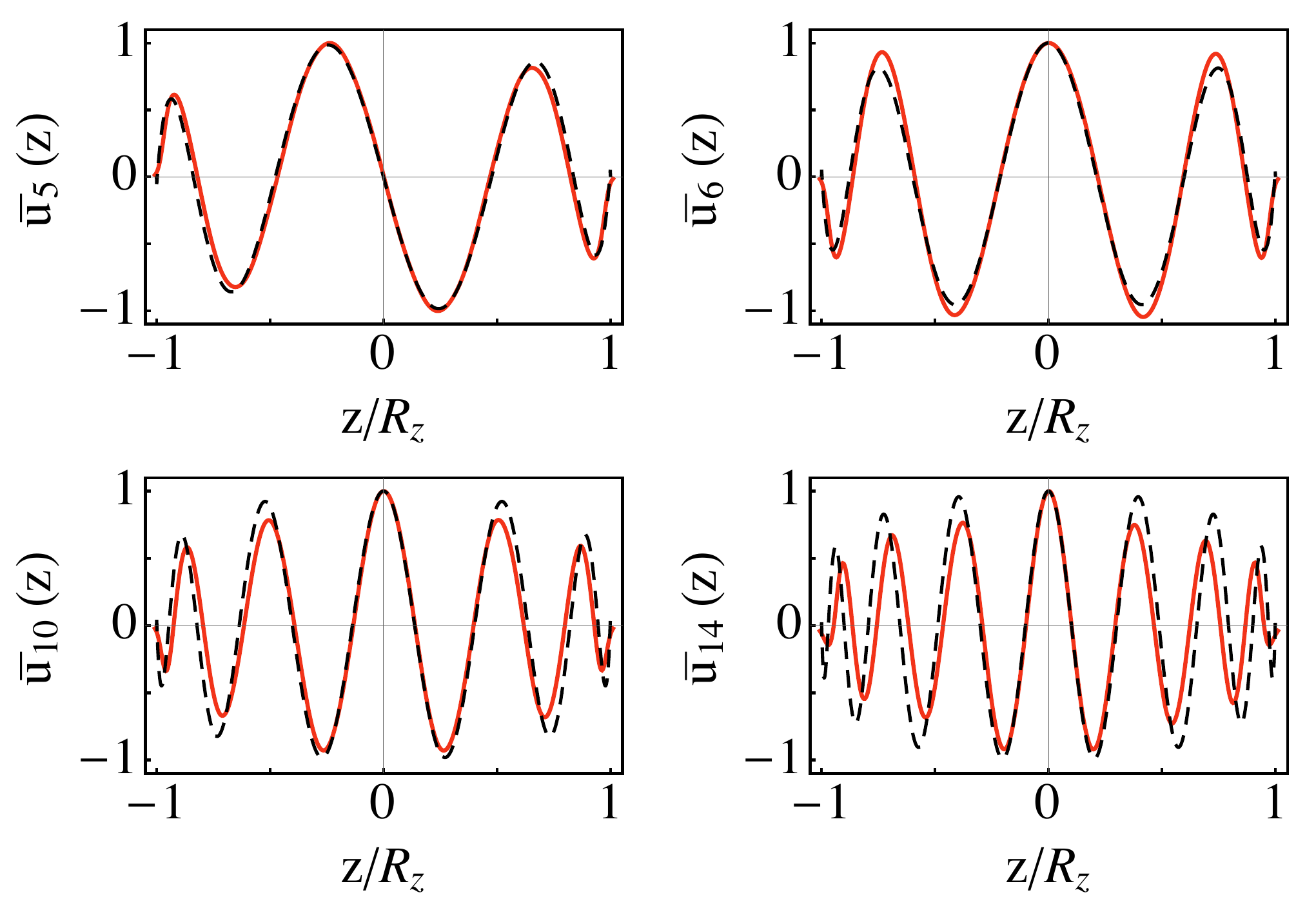}}
\caption{Analysis of the eigenmodes. Upper panel:  Comparison between PCA-extracted mode frequencies and the theoretical eigenmodes of an elongated Bose-Einstein condensate (dashed line). Lower panels: space  structure of modes  $k=$5, 6, 10, and 14, and their fit using Geigenbauer polynomials.}
\label{fig:PCAvsWKB}
\end{figure}

We monitor the mode coupling by plotting the evolution of the weight $c_{ik}(t)$ of the $k$-th mode of the superfluid $i$ unveiled by the PCA (see Fig. \ref{fig:param}).   As a first check, we plot the Fourier spectrum of $c_{ik}$ to extract the frequencies of each mode. In Fig. (\ref{fig:PCAvsWKB}) we compare the PCA-extracted oscillations with the analytic prediction of frequencies in elongated traps \cite{stringari1998dynamics}. For $\mu\gg \hbar\omega_\perp$ the cloud is hydrodynamic in all three directions and the frequency of the $k$-th mode is:
\be\omega_k=\frac{1}{2}\sqrt{k (k+3)}\omega_z.
\label{Eq:Hydro3D}
\ee
This assumption is satisfied by the larger cloud, as illustrated in Fig. \ref{fig:PCAvsWKB} by the agreement between the numerical results and the predictions of Eq. (\ref{Eq:Hydro3D}). By contrast, 3D hydrodynamics breaks down for the smaller cloud where the low atom number yields a lower chemical potential.  In this regime,  since $\mu\lesssim\hbar\omega_\perp$, the transverse degrees of freedom are frozen and the collective dynamics is one-dimensional, leading to the following dispersion relation
\be\omega_k=\sqrt{\frac{k (k+3)}{2}}\omega_z,
\label{Eq:Hydro1D}
\ee
which is in very good agreement with the results of numerical simulations (see the upper panel of Fig. \ref{fig:PCAvsWKB}) .

Likewise, we compare in the lower panel of Fig. \ref{fig:PCAvsWKB} the spatial structure of the modes unveiled by the PCA and show that they agree with the Geigenbauer polynomials known to describe the low-lying modes of a harmonically trapped  Bose-Einstein condensate \cite{jin2019hydrodynamics}.

\section{Hydrodynamic approach}

\subsection{General formalism}

To explain  the excitation mechanism of the modes observed in numerical simulations, we analyze their dynamics using  the  hydrodynamic approximation. This approach is rather standard and is summarized in the following section. The starting point of this second analysis is the classical Hamiltonian
\be
\begin{split}
H=\int d^3\bm r \Big[\sum_{i=1}^N&\left(\frac{\hbar^2}{2m_i}\rho_i(\bm r)\left(\nabla\Phi_i\right)^2+\rho_i U_i(\bm r)\right)\\
&+e(\rho_1,...,\rho_N)\Big],
\end{split}
\ee
that describes the dynamics of an ensemble of $N$ superfluids. In this expression, $\Phi_i$ and $\rho_i$ are the phase and  the density of superfluid $i$, while $e$ is the energy density of the system. Taking the phase and density as dynamically conjugate variables, Hamilton's equations of motion yield
\begin{eqnarray}
m_i\left(\partial_t\bm v_i+\nabla \bm v_i^2/2\right)=-\nabla\left(\frac{\partial e}{\partial\rho_i}+U_i\right)\\
\partial_t\rho_i+\nabla\left(\rho_i\bm v_i\right)=0,
\end{eqnarray}
where $\bm v_i=\hbar\nabla\Phi_i/m_i$  is the local superfluid velocity and $\partial_{\rho_i}e$ is the chemical potential of species $i$. In the limit of weak interspecies-coupling, we can expand the chemical potential versus the densities of the other components:
$$
\frac{\partial e}{\partial\rho_i}\simeq\mu_i(\rho_i)+\sum_{j\not = i}g_{ij}\rho_j,
$$
where $\mu_i(\rho_i)$ is species $i$ alone's zero-temperature equation of state. Since by definition we have $g_{ij}=\partial^2_{\rho_j\rho_i}e$, the interspecies coupling constants obey the symmetry relation $g_{ij}=g_{ji}$.

Consider first a single species ($g_{ij}=0$). In the stationary state,  the phase varies as $\Phi_{i,0}=\mu_i^0 t/\hbar$ and the density profile $\rho_{i,0}(\bm r)$ is time independent and satisfies a Thomas-Fermi equation
\be
\mu_i(\rho_{i,0}(\bm r))+U_i(\bm r)=\mu_i^0,
\label{EqLDA}
\ee
 where the index $0$ indicates equilibrium quantities.

Low-lying excitations are obtained by considering $\bm v_i$ and $\delta\rho_i=\rho_i-\rho_{i,0}$ as small parameters. Expanding the hydrodynamic equation yields after some straightforward algebra
\be
\partial_t^2\delta\mu_i-\frac{1}{m_i}\left(\frac{\partial\mu_i}{\partial\rho_i}\right)_0\nabla\left(\rho_{i,0}\nabla\delta\mu_i\right)=0.
\ee
with $\delta\mu_i=\left(\partial\mu_i/\partial\rho_i\right)_0\delta\rho_i$.

The eigenmodes of the superfluid are found using the ansatz $\delta\mu_i(\bm r,t)=u_{i,k}(\bm r)e^{-i\omega_{i,k}t}$, where $u_{i,k}$ and $\omega_{i,k}$ are solutions of the eigenproblem
\be
\omega_{i,k}^2u_{i,k}={\cal L}_i[u_{i,k}],
\label{EqEigen}
\ee
with
\be
{\cal L}_i[u]=-\frac{1}{m_i}\left(\frac{\partial\mu_i}{\partial\rho_i}\right)_0\nabla\left[\rho_{i,0}\nabla u\right].
\ee
${\cal L}_i$ is a positive operator and is symmetric for the inner product $\langle\cdot|\cdot\rangle_i$ defined by
\be
\langle u| v\rangle_i=\int d^3\bm r \left(\frac{\partial \rho_i}{\partial\mu_i}\right)_0u(\bm r)^*v(\bm r).
\label{scalar-product}
\ee
The $u_{i,k}$ can then be chosen as an orthonormal basis for this inner product.

In the literature, general properties of Eq. (\ref{EqEigen}) were derived for simple cases:
\begin{itemize}
\item $u_i$=cte is a solution to Eq. (\ref{EqEigen}) for the eigenfrequency $\omega_i=0$. This solution corresponds to a  shift of the global chemical potential $\mu_i^{0}$ in Eq. (\ref{EqLDA}) and implies that all other eigenstates $u_{ik}$ are orthogonal to constant functions for the inner product  (\ref{scalar-product}).

\item For a homogenous system, the solutions to Eq. (\ref{EqEigen}) are plane waves and describe first-sound propagation. The eigenfunctions are characterized by a wave-vector $\bm k_i$ and obey the dispersion delation $\omega_i=c_i k_i$, where the sound velocity is defined by $m_i c_i^2=\rho_{i,0}(\partial\mu_i/\partial\rho_i)_0$.
\item The plane-wave structure also applies to cylindrical traps \cite{stringari1998dynamics}. In this case, the waves propagate along the symmetry axis of the cloud at a velocity $c_{1D}=\sqrt{\bar\rho_{i,0}} (\partial\mu_i/\partial\bar\rho_i)_0/m_i]$, where $\bar\rho_i$ is  the density integrated over the transverse degrees of freedom.
\item In a harmonic trap, we recover Kohn's theorem \cite{kohn1961cyclotron} \Blue{since the oscillation of the center of mass along the principal direction $x_\alpha$ of the trap corresponds} to the function $u_{ik}=x_\alpha$ and are associated with the eigenvalue $\omega_{i,\alpha}$.
\item The previous result can be generalized to any mode for polytropic equation of state $\mu_i\propto\rho_i^\gamma$. In this case,  the eigenmodes of ${\cal L}_i$ are Gegenbauer polynomials of the spatial coordinates.
\end{itemize}

Furthermore, for a harmonic (or flat) potential, the dynamics of the cloud obeys Kohn's theorem \cite{kohn1961cyclotron} and the previous results can be extended to the case where the superfluid is initially oscillating. More precisely, the hydrodynamic equations are invariant under the transformation $\bm r\rightarrow \bm r'=\bm r-\bm R_i(t)$, $\bm v_i\rightarrow\bm v'_i=\bm v_i-\bm V_i$, where $\bm R_i$ and $\bm V_i$ are the position and the velocity of the center of mass of the cloud and obey Newton's second law $m_i\dot{\bm V_i}=-\nabla U_i(\bm R_i)$. As a consequence, the eigenmodes $u'_{i,k}(\bm r,t)=u_{i,k}(\bm r-\bm R_i(t))$ can also be used to describe the low-lying excitations of an oscillating superfluid \footnote{This result also applies in the special case  $U_i=0$,  where the center of mass moves at a constant velocity and in which case, Kohn's theorem coincides with Galilean invariance.}.

We now consider an ensemble of coupled superfluids. As before, in absence of coupling, the density in the laboratory frame is given by $\rho_i(\bm r,t)=\rho_{i,0}(\bm r-\bm R_i(t))$. The coupling affects the density profiles which can be expanded over the eigenmodes $u_{i,k}$ as
\be
\begin{split}
\rho_i(\bm r,t)&=\rho_{i,0}(\bm r-\bm R_i(t))\\
&+\sum_k c_{i,k}(t)\left(\frac{\partial\rho_i}{\partial\mu_i}\right)_0(\bm r-\bm R_i(t))u_{k,i}(\bm r-\bm R_i(t)).
\end{split}
\label{EqnExpansion}
\ee

To simplify the notation, we hereafter denote with a prime physical quantities evaluated in the moving frame. In other words, for any function $F_i(\bm r)$, we define $F'_i(\bm r,t)$ as $F'_i(\bm r,t)=F_i(\bm r-\bm R_i(t))$.

Inserting this expansion in the hydrodynamic equations, we obtain the following set of coupled differential equations for the coefficients $c_{i,k}$
\be
\begin{split}
\ddot c_{i,k}&+\omega_{i,k}^2c_{i,k}\\
&+\omega_{i,k}^2\sum_{j\not = i}g_{ij}\left[A_{i,j,k}(t)+\sum_{k'}B_{ik,jk'}(t)c_{jk'}\right]=0,
\end{split}
\label{EqCoupledModes}
\ee
where
\begin{eqnarray}
A_{ijk}(t)&=&\langle u'_{i,k}(\bm r,t)|\rho'_{j,0}(\bm r,t)\rangle_i\label{EqnCoupled}\\
B_{ik,jk'}(t)&=&\langle u'_{i,k}(\bm r,t)|\left(\frac{\partial\rho_j}{\partial\mu_j}\right)'_0 u'_{j,k'}(\bm r,t)\rangle_i
\end{eqnarray}
are time-dependent coefficients (note that by definition of the inner product, $B_{ik,jk'}=B_{jk',ik}^*$).

\subsection{Homogeneous cloud}

We first consider the special case of a homogeneous system. The static density profiles are uniform, and the eigenmodes $u_{jk}$ are therefore orthogonal to $\rho_{j'}$. Coefficients $A_{ijk}$ vanish and the dynamics is set by the coupling coefficients $B_{ik,jk'}$.

As stated earlier, the unperturbed superfluids move at constant velocities $\bm V_i$ and the eigenmodes  are plane waves characterized by  wave-vectors $\bm k_{j=1..N}$. We then have
\be
B_{ik, jk'}(t)=g_{i,j} \sqrt{\frac{\partial\rho_i}{\partial\mu_i}\frac{\partial\rho_j}{\partial\mu_j}} e^{i \bm k_i (\bm V_i-\bm V_j)t}\delta_{\bm k_i,\bm k_j}.
\ee

Under these assumptions, the linearized hydrodynamic equations (\ref{EqCoupledModes}) can be written as
\be
\ddot c_{i,\bm k}+\omega_{i,\bm k}^2c_{1,\bm k}=-\omega_{i,\bm k}^2 \sum_{j\not = i}g_{ij} \sqrt{\frac{\partial\rho_i}{\partial\mu_i}\frac{\partial\rho_j}{\partial\mu_j}} c_{j,\bm k} e^{i\bm k\cdot \bm V_{ij}t}\label{EqParam0}
\ee

From (\ref{EqParam0}), we infer that the inter-species interactions only couple same-momentum modes and that the coupling constant oscillates in time at a frequency $\bm k \cdot (\bm V_i-\bm V_j)$. This behaviour is reminiscent of a parametric oscillator characterized by a dynamical instability that we interpret as follows: In absence of coupling, the free solutions of Eq. (\ref{EqParam0}) are $c_{i,\bm k}\propto \exp(\pm i\omega_{i,\bm k}t)$. To the leading order of the perturbation, we can insert this solution in the right-hand side of Eq. (\ref{EqParam0}) which is now equivalent to an ensemble of $N$ harmonic oscillators driven at frequencies $\bm k\cdot\bm V_{ij}\pm \omega_{j,\bm k}$. We  notice that when
\be|\bm k\cdot \bm V_{ij}|\simeq \omega_{i,\bm k}+\omega_{j,\bm k},\label{Eq:Instability}\ee
 the drive is resonant and leads to the instability of the system. We prove the existence of this instability in the the appendix \ref{AppendixParam} for a mixture of $N=2$  superfluids.

For 3D phonons characterized by a velocity $c_j$, excitations can propagate in any direction and the instability criterion (\ref{Eq:Instability}) can be reformulated as $V_{ij}\ge c_i+c_j$, which is the extended Landau criterion \cite{castin2014landau}. For an elongated harmonic trap, this argument is still valid but phonons can only propagate along the trap axis. In this case the instability criterion selects a velocity window centered on $c_i+c_j$ (see for instance \cite{abad2014counter} and Appendix \ref{AppendixParam}).

\subsection{Trapped superfluids}

We consider now the case of a harmonically trapped mixture of superfluids. The parametric mechanism discussed in the case of a homogeneous system is still present. Consider for simplicity a situation where the trapping frequencies of the two clouds are equal. In this case, the relative position and the coefficients $B_{ik,jk'}$ oscillate with frequency $\omega_z$. After expanding these coefficients, the qualitative argument put forward for homogeneous systems predicts that the parametric amplification of pairs of  spatially-matching modes occurs when their frequencies meet the resonance conditions
\be
\omega_{ 1,\bm k}+\omega_{2,\bm k}=n\omega_z,
\label{resonance}
\ee
where $n$ is an integer. In Apppendix \ref{AppendixToy} we present a toy model leading to this resonance condition. This parametric mechanism is illustrated in the right panel of Fig.  \ref{fig:param} where we notice that the modes excited in the two clouds share approximately the same wavelength.

Another specific feature of the trap is the inhomogeneity of the density profiles allowing for non-zero coupling coefficients $A_{ijk}$.  Qualitatively, this mechanism can be interpreted as an excitation of the superfluid $i$ by the mean-field potential $g_{ij}\rho_{j,0}(\bm r-\bm R_j (t))$. Moreover, this coupling affects only the large cloud. Indeed, at the scale of the small cloud $j$, the density profile of a larger cloud $i$ is approximately flat and the overlap between $\rho_{i,0}$ and any mode $u_{jk}$ is therefore vanishingly small according to the general properties of the operator $\cal L$.

To interpret the result of the simulation displayed in Fig. \ref{fig:param}, we consider the case $N=2$, with, for simplicity, $B_{ik,jk'}=0$.  We note that Eq. (\ref{EqCoupledModes}) is formally  equivalent to that of a driven linear oscillator. Moreover, if we expand the driving term with the relative displacement $z_{ij}=(\bm R_i(t)-\bm R_j(t))\cdot \bm u_z$, we obtain \footnote{Note that for a mean-field BEC in the Thomas-Fermi regime, the density profile is quadratic in position and the sum ends at $n=3$.}
\be
A_{ij,k}=\sum_{n=1}^\infty (-1)^nz_{ij}^n \langle u_i|\partial_z^n\rho_j\rangle_i.
\ee

\begin{figure}
\includegraphics[width=\columnwidth]{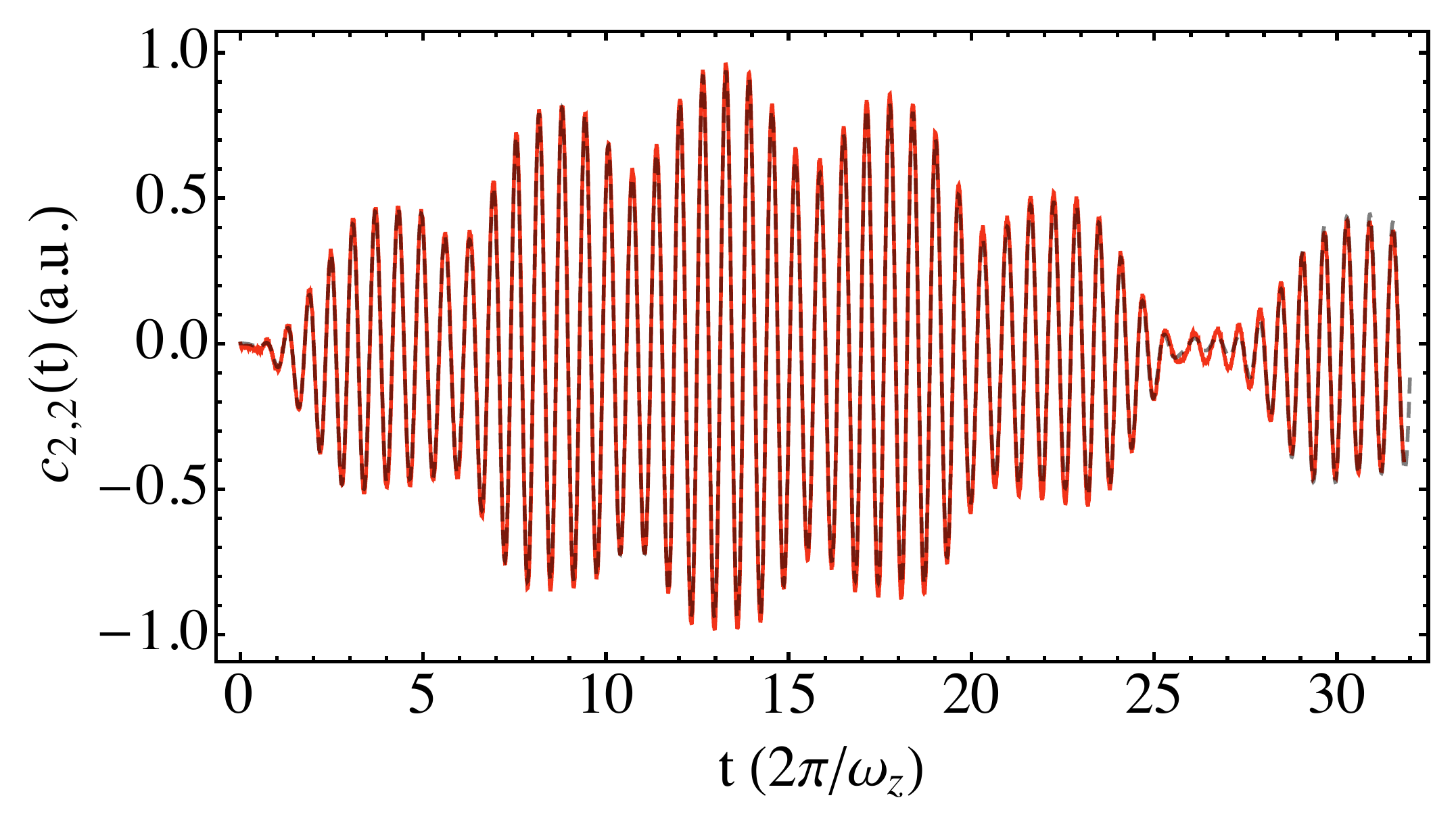}
\caption{Spectral weight of the breathing mode ($k=2$) of the large cloud. Numerical results (blue solid line) are fitted using a driven oscillator model (black dashed line).}
\label{Fig1}
\end{figure}

In Fig. (\ref{Fig1}), we assess the accuracy of this model using the first two harmonics ($n=1$ and $n=2$) of the dynamics of the breathing mode ($k=2$) obtained from numerical simulations. We consider the amplitudes and the frequencies of the breathing mode as free parameters and we find $\omega_{2,2}=1.580(3)\omega_z$, in agreement with the expected value $\omega_{2,2}/\omega_z=\sqrt{5/2}\simeq 1.581$.

Finally, we note that in the case of the dipole mode $k=1$, the excitation is resonant with the mode-frequency and triggers a slow divergence of the amplitude of the mode. In Appendix \ref{AppendixCdM} we show that this resonant behaviour gives rise to the dipole-mode frequency shift,  expressed in Eq. \ref{EqnDipoleShift}.

\section{Conclusion and implications for the superfluid critical velocity in trapped systems}

Both numerical and analytical approaches conclude to the existence of two mechanisms for the coupling of the center of mass motion to low-lying excitation modes. The parametric scenario is associated with  the formation of pairs of excitations in both superfluids. By contrast, the linear mechanism can be interpreted as the resonant creation of excitation in the large cloud. This mechanism is the analogous to the motion of a potential in the larger superfluid and is consequently very similar to Landau's argument. Our zero-temperature approach predicts a coherent mode-coupling. At finite temperature, these modes are damped and will pump energy out of motion of the center of mass of the two clouds, as observed experimentally \cite{wen2017dipole,Lee16Phase,Edmonds15Nonequilibrium}.

The resonance condition for the parametric instability can be reinterpreted in terms of relative velocity of the two clouds. By analogy with the homogeneous case, the instability is triggered when the relative velocity cross the instability window centered on $c_1+c_2$. This scenario is supported by the toy model presented in Appendix \ref{AppendixToy} and is validated by our numerical simulations (see Fig. \ref{fig:param}, where we observe that the amplitude of the mode grows only inside narrow time windows). By contrast, the linear coupling predicts a critical velocity equal to the sound velocity of the large cloud. In an imbalanced system as in \cite{delehaye2015critical}, $c_1\ll c_2$ and the two velocities are similar. Experimentally, the critical velocity reported in \cite{delehaye2015critical} is therefore compatible with  both scenarios.

\acknowledgments
The authors thank N. Proukakis, C. Salomon and ENS ultracold Fermi group for helpful discussions. F. Chevy acknowledges support from ANR (grant SpifBox), ERC (advanced grant CritiSup2) and Fondation del Duca. P. Parnaudeau and I. Danaila acknowledge financial support from the French ANR grant ANR-18-CE46-0013 QUTE-HPC. Part of this work used computational resources provided by IDRIS (Institut du d{\'e}veloppement et des ressources en informatique scientifique) and CRIANN (Centre R{\'e}gional Informatique et d'Applications Num{\'e}riques de Normandie).

\appendix

\section{Details of numerical simulations}
\label{AppendixSim}

\begin{figure}[!h]
	\includegraphics[width=\columnwidth]{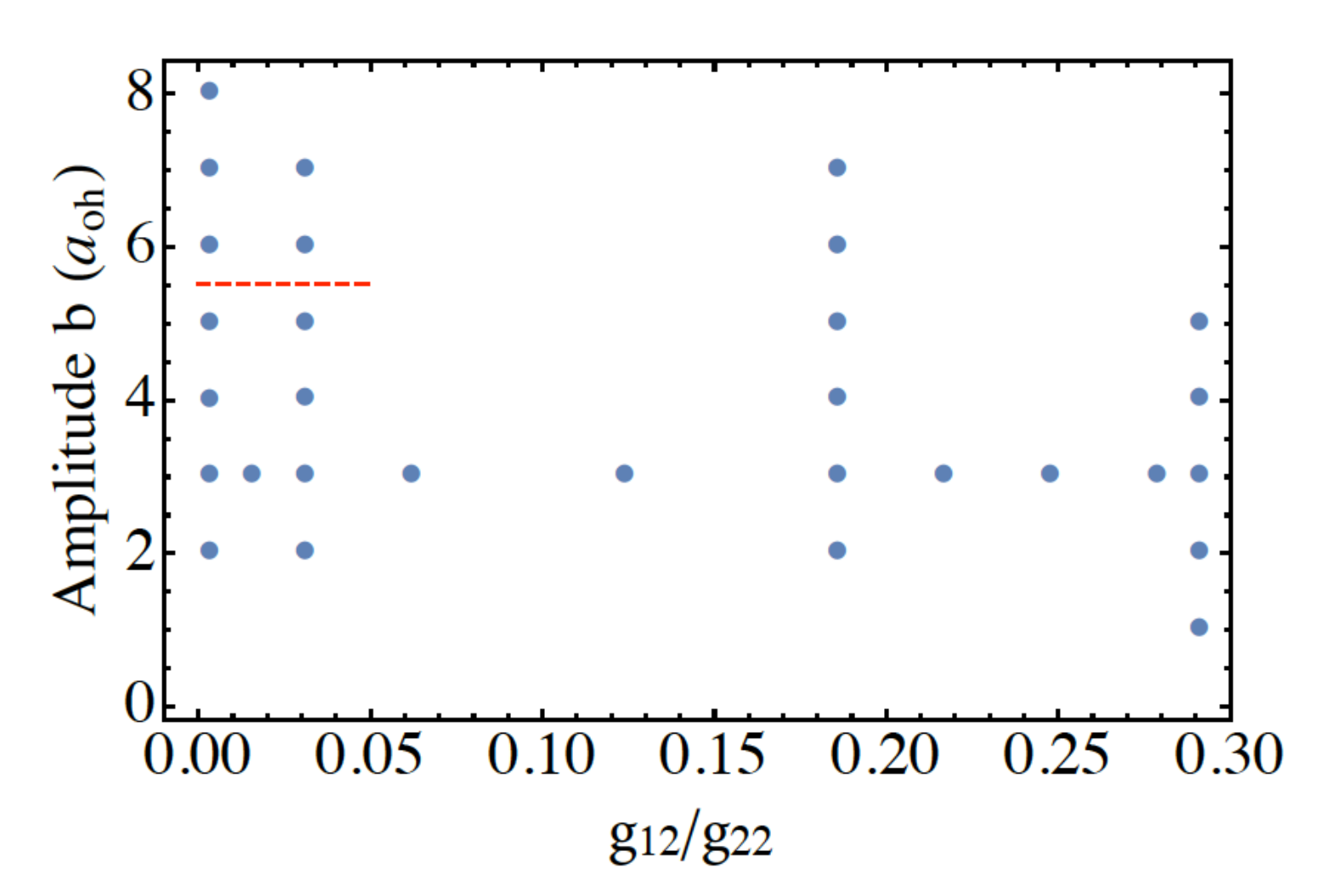}
	\caption{Summary of parameters for which simulations were performed (blue dots).
The red dashed line shows the amplitude above which the critical velocity is expected to
be reached in the limit of a vanishing coupling between the superfluids.}	
	\label{fig:Runs}
\end{figure}

Numerical simulations are performed using dimensionless variables. We take the usual scaling:
\begin{equation}
{\bf x} \rightarrow \frac{\bf x}{x_s}, \quad {t} \rightarrow \frac{t}{t_s}, \quad
u_1 = \frac{{\psi_1}}{ x_s^{-3/2}}, \quad u_2 = \frac{{\psi_2}}{ x_s^{-3/2}}.
\label{eq-GPn_scal_var}
\end{equation}
with
\begin{equation}
t_s =\frac{1}{\omega}, \quad x_s = a_{ho}, \,\, a_{ho}=\sqrt{\frac{\hbar}{m_2 \omega_z}}.
\label{eq-GPn-scales}
\end{equation}
The  non-dimensional form of Eqs.  (\ref{eq-GP_psi1})-(\ref{eq-GP_psi2}) becomes:
\begin{eqnarray}
\label{eq-GPn_psi1}
\ds i \frac{\partial u_1}{\partial t} &=&\ds  \left[-\frac{\nabla^2}{2 d_1}   + U_a({\bf r})+ \beta_{11} |u_1|^2 + \beta_{12} |u_2|^2\right] u_1,\\
\label{eq-GPn_psi2}
\ds i \frac{\partial u_2}{\partial t} &=&\ds  \left[-\frac{\nabla^2}{2 d_2}   + U_a({\bf r}) + \beta_{21} |u_1|^2 + \beta_{22} |u_2|^2 \right]u_2,
\end{eqnarray}
with $u_1$ and $u_2$ normalized to unity:
\begin{equation}
\int_{R^3} |u_1|^2 ={1}, \quad \int_{R^3} |u_2|^2 ={1}.
\label{eq-GPn-unity}
\end{equation}
The non-dimensional trapping potential takes into account the initial shift $b$ (in $a_{ho}$ units) of the clouds:
\begin{equation}
U_a({\bf r})= \frac{d_2}{2} \left[ \gamma_\perp^2 ({x}^2 +  {y}^2) + \gamma_z^2 \left(z-b(t)\right)^2\right],
\label{eq-GPn-pot}
\end{equation}
where $\gamma_\perp=\left({\omega_\perp}/{\omega}\right), \quad \gamma_z=\left({\omega_z}/{\omega}\right)$ and $b(t)=0$ for $t>0$.
Dimensionless parameters in Eqs.  (\ref{eq-GPn_psi1})-(\ref{eq-GPn_psi2}) are expressed by:
\begin{eqnarray}
\beta_{11} = {\displaystyle 4\pi \frac{1}{d_1} \frac{N_1 a_{11}}{a_{ho}}},& \beta_{12} = \displaystyle {\displaystyle 2\pi \frac{d_1+d_2}{d_1 d_2} \frac{N_2 a_{12}}{a_{ho}}},\\
\beta_{22} = \displaystyle {\displaystyle 4\pi \frac{1}{d_2} \frac{N_2 a_{22}}{a_{ho}}}, & \beta_{21} = \displaystyle {\displaystyle 2\pi \frac{d_1+d_2}{d_1 d_2} \frac{N_1 a_{12}}{a_{ho}}}.
\end{eqnarray}
Each run is identified by the value of the parameter $g_{12}/g_{22} = \beta_{12}/\beta_{22}$ and the value of $b$, the initial shift of the clouds. The explored values of these parameters is depicted in Fig. \ref{fig:Runs}. The long-time integration for these  30 cases requested over 50000 hours of CPU time and generated teraoctets of data.

To avoid  clouds coming close to the boundaries during oscillations, the dimensions of the computational domain
were fixed as $L_x = L_y = 8\, a_{ho}$ and $L_z = 64 \, a_{ho}$.
This allowed us to capture amplitude oscillations up to $16 \, a_{ho}$ along the $z$ direction,
which is more than enough to obtain a relative velocity larger than the sum of the central sound
velocities of the condensates (typically reached for  $5.5 \, a_{ho}$).
The grid resolution was $\delta x= \delta y = \delta z = a_{ho}/16$, resulting in computational grids of
$128\times128\times1024$ points following $x$, $y$ and $z$ directions, respectively.

To capture both fast and slow dynamics present during the oscillations of the
clouds, we used a refined time resolution $\delta t = 5\cdot10^{-4}/\omega_z$  and a large number of steps
$N_t = 4\cdot 10^5$. This allowed us to simulate  at least 32 periods of oscillations for each run case.

\section{Parametric instability criterion for a mixture of two homogeneous superfluids}
\label{AppendixParam}

 We prove here the existence of parametric instability for  a mixture of $N=2$ superfluids. In this case, Eq. (\ref{EqParam0}) reduces to a set of decoupled two-dimensional problems governed by the following equations

\begin{eqnarray}
\ddot c_{1,\bm k}+\omega_{1,\bm k}^2c_{1,\bm k}&=&-\omega_{1,\bm k}^2 g_{12} \sqrt{\frac{\partial\rho_1}{\partial\mu_1}\frac{\partial\rho_2}{\partial\mu_2}} c_{2,\bm k} e^{i\bm k\cdot \bm V_{12}t}\label{EqParam1}\\
\ddot c_{2,\bm k}+\omega_{2,\bm k}^2c_{2,\bm k}&=&-\omega_{2,\bm k}^2   g_{12} \sqrt{\frac{\partial\rho_1}{\partial\mu_1}\frac{\partial\rho_2}{\partial\mu_2}}c_{1,\bm k} e^{-i\bm k\cdot \bm V_{12}t}\label{EqParam2}
\end{eqnarray}
with $\bm V_{1,2}=\bm V_1-\bm V_2$. These equations are  solved by taking  $c_{1,\bm k}=\tilde c_{1,\bm k}^{(0)}\exp[(\gamma_{\bm k}+i\bm k\cdot \bm V_{12}/2)t]$ and $c_{2,\bm k}=\tilde c_{2,\bm k}(t)\exp[(\gamma_{\bm k}-i\bm k\cdot \bm V_{12}/2)t]$.
The amplification rate $\gamma$ is thus solution to the eigenvalue equation
\be
\begin{split}
\left[\left(\gamma-i\frac{\bm k\cdot\bm V_{12}}{2}\right)^2+\omega_{1,\bm k}^2\right]\left[\left(\gamma+i\frac{\bm k\cdot\bm V_{12}}{2}\right)^2+\omega_{2,\bm k}^2\right]\\
=g_{12}^2\omega^2_{1,\bm k}\omega^2_{2,\bm k}\left(\frac{\partial\rho_1}{\partial\mu_1}\right)\left(\frac{\partial\rho_2}{\partial\mu_2}\right)
\end{split}
\ee

We recover the equations derived in \cite{abad2014counter} showing that for a small coupling, the acoustic modes become dynamically unstable when the velocity satisfies the condition
\be
|\bm V\cdot\bm u-V_c|\le g_{12}\sqrt{\frac{\omega_{1,\bm k}\omega_{2,\bm k}}{\frac{\partial\rho_1}{\partial\mu_1}\frac{\partial\rho_2}{\partial\mu_2}}},
\ee
with $V_c=c_1+c_2$ and $\bm u=\bm k/k$.

\section{A toy model for the parametric instability in a trap}
\label{AppendixToy}

We describe the parametric  mode-coupling in a trap using the following assumptions. First, we approximate the space structure of the mode by local plane waves characterized by a wave-vector $k=\omega_k/c_i$  (this is true for high frequency mode at the center of the trap, following the WKB approximation \cite{jin2019hydrodynamics}). Second, we assume that the trap frequencies of the two clouds are identical and that the initial displacement of the two clouds are different. As a consequence, the relative distance between the centers of the two clouds evolves as $R\cos(\Omega t)$, where $\Omega$ is the common trapping frequency.

With these assumptions, the amplitudes of two matching modes will satisfy equations similar to Eqs.  (\ref{EqParam1})-(\ref{EqParam2}), where the relative displacement $\bm V_{12}t$ is replaced by $R\cos(\Omega t)$.
\begin{eqnarray}
\ddot c_{1,\bm k}+\omega_{1,\bm k}^2c_{1,\bm k}&=&\omega_{1,\bm k}^2 \varepsilon c_{2,\bm k} e^{ikR\cos(\Omega t)}\\
\ddot c_{2,\bm k}+\omega_{2,\bm k}^2c_{2,\bm k}&=&\omega_{2,\bm k}^2    \varepsilon c_{1,\bm k} e^{ikR\cos(\Omega t)}.
\end{eqnarray}

We can repeat the qualitative argument developed for the homogeneous cloud by noting that $\exp(ikR\cos(\Omega t))$ can be expanded as
\be
e^{ikR\cos(\Omega t)}=\sum_n i^n J_n(kR)e^{n\Omega t},
\label{Eq:Bessel}
\ee
where $J_n$ are Bessel functions of the first kind. As before, each harmonic of the sum will give rise to a parametric instability when the resonance condition $n\Omega=\omega_{1,k}+\omega_{2,k}$ is met and, in this case, the effective coupling is proportional to $J_n(kR)$. Consider now the case of a coupling with a high-frequency phonon mode, as in Fig. \ref{fig:param}). In this case, $\Omega\ll \omega_{\alpha,k}$ and $n\gg1$. But, at a given $kR$, $J_n(kR)$ is vanishingly small for large $n$ and,  to maintain a significant coupling, we need to take $kR\gtrsim n$. Indeed, from Eq. (\ref{Eq:Bessel}), we infer that
$$
J_n(kR)=\frac{1}{2i^n\pi}\int_0^{2\pi}d\theta e^{-i(n\theta+kR\cos\theta)}
$$
For large values of $n$, the exponential term oscillates rapidly, unless the phase is stationary. The integral will therefore be dominated by values of $\theta$ close to $\theta^*$, defined by the stationary phase condition $kR\sin\theta^*=n$, implying that $kR\ge n$. Since $n\Omega=\omega_{k,1}+\omega_{k,2}$, this condition can be recast as

\be
\Omega R\ge \frac{\omega_{k,1}+\omega_{k,2}}{k}=c_1+c_2
\ee
We therefore recover that, even for an oscillatory motion, dissipation occurs when the maximal relative velocity $\Omega R$ is above the sum of the sound velocities of the two superfluids.

\section{Dipole mode frequency shift}
\label{AppendixCdM}
We consider here the linear coupling to the  mode  $k=1$ corresponding to the dipole mode along the direction $z$. As mentioned in the general properties of the linearized hydrodynamic equations, this mode is  associated with the eigenvector  $u_{i,1}=z/\|z\|_i$, with  eigenfrequency $\omega_{i,1}=\omega_{z,i}$. We assume that the larger superfluid ($i=2$) is at rest while the smaller one ($i=1$) oscillates with an amplitude $A$.

The time-evolution of the dipole-mode amplitude $c_{1,1}$ is therefore driven by the coupling coefficient
\begin{eqnarray}\nonumber
& &A_{12,1}=\int d^3\bm r\left(\frac{\partial\rho_1}{\partial\mu_1}\right)_0\rho_{2,0}(\bm r+A\cos(\omega_{z,1} t)\bm u_z)\frac{z}{\|z\|_1}\\
&\simeq&-m_2\omega_{z,2}^2 A\cos(\omega_{z,1} t)\int d^3\bm r\left(\frac{\partial\rho_2}{\partial\mu_2}\right)_0\left(\frac{\partial\rho_1}{\partial\mu_1}\right)\frac{z^2}{\|z\|_1},
\end{eqnarray}
where we assumed that the amplitude $A$ of the motion is smaller than the size of the larger ($i=2$) cloud. Using the fact that the size of the static cloud is larger than that of the moving one and that the trapping potentials are identical for the two species, we can further simplify this expression as
\be
A_{12,1}\simeq -m_1\omega_{z,1}^2 A\left(\frac{\partial\rho_2}{\partial\mu_2}\right)_0(\bm r=0)\|z\|_1\cos(\omega_{z,1} t)
\ee
We note that the driving term $A_{12,1}$ oscillates at the resonance frequency $\omega_{z,1}$ of the dipole mode. The solution to the equation
\be
\ddot c_{11}+\omega_{z,1}^2 c_{11}=-g_{12}\omega_{z,1}^2 A_{12,1}
\ee
is therefore characterized by a linearly divergent behaviour in long time, and we obtain in the steady state regime
\be
c_{11}(t)\sim \frac{g_{12}}{2} m_1\omega_{z,1}^3 A\left(\frac{\partial\rho_2}{\partial\mu_2}\right)_0(\bm r=0)\|z\|_1 t\sin(\omega_{z,1} t).
\label{EqnDipole}
\ee

The onset of this divergence in a perturbation expansion is usually the signature of a shift of the natural oscillation frequency of the system. Let us indeed assume that the oscillation is shifted by $\delta\Omega$. The density profile of species $i=1$ now evolves as
$\rho_1(\bm r,t)=\rho_{1,0}(z-A\cos((\omega_{z,1}+\delta\Omega)t))$. Using the fact that in the stationary regime, the density profile of the unperturbed system is given by the LDA expression $\rho_{1,0}(\bm r)=\rho_{1}(\mu_1^{(0)}-V(\bm r))$, expanding the density profile to first order in $\delta\Omega$ yields
\be
\begin{split}
  \rho_1(\bm r,t)&=\rho_{2,0}(z-A\cos(\omega_{z,1} t))\\
  &-\left(\frac{d\rho_1}{d\mu_1}\right)_0m_1\omega_{z,1}^2 z A\delta\Omega t\sin(\omega_{z,1} t)
  \label{eqsplit}
\end{split}
\ee
We recover the frequency shift expressed in Eq. (\ref{EqnDipoleShift}) by comparing Eq. (\ref{eqsplit}) to the general expansion (\ref{EqnExpansion}), where $c_{1,1}$ is given by  Eq. (\ref{EqnExpansion}).

\bibliographystyle{unsrt}
\bibliography{bibliographie}

\end{document}